\documentclass[aps,twocolumn,showpacs]{revtex4}
\usepackage{amsmath}
\usepackage{epsfig}
\usepackage{booktabs} %调整表格线与上下内容的间隔
\usepackage{multirow}

\begin{document}

\noindent\makebox[\textwidth]{\footnotesize To be published in Chinese Physics C}

\title{Investigation of fully heavy tetraquark within chiral quark model}

\author{Yuheng Wu$^1$}%\email[E-mail: ]{191002007@njnu.edu.cn}
\author{Xuejie Liu$^2$}%\email[E-mail: ]{1830592517@qq.com}
\author{Ye Yan$^3$}
\author{Yue Tan$^1$}%\email[E-mail: ]{tanyue@ycit.edu.cn }%(Corresponding author)}
\author{Qi Huang$^4$}\email[E-mail: ]{06289@njnu.edu.cn}
\author{Hongxia Huang$^4$}\email[E-mail: ]{hxhuang@njnu.edu.cn} %(Corresponding author)}
\author{Jialun Ping$^4$}%\email[E-mail: ]{jlping@njnu.edu.cn}

\affiliation{$^1$Department of Physics, Yancheng Institute of Technology, Yancheng 224000, P. R. China}
\affiliation{$^2$Department of Physics, Henan Normal University, Xinxiang, Henan 453007, P. R. China}
\affiliation{$^3$Department of Physics, Changzhou College of Information Technology, Changzhou, Jiangsu 213164, P. R. China}
\affiliation{$^4$Department of Physics, Nanjing Normal University, Nanjing, Jiangsu 210097, P. R. China}

\begin{abstract}
In the framework of the Chiral quark model (ChQM), we investigate the fully charmed and fully bottomed tetraquark with $J^{PC}=2^{++}$ including two structures: $Q\bar{Q}-Q\bar{Q}$ and $QQ-\bar{Q}\bar{Q}$. The bound-state calculation shows that there is no bound state in either $cc\bar{c}\bar{c}$ or $bb\bar{b}\bar{b}$ systems.
However, by using the real-scaling method, some resonance states are obtained.
For the $cc\bar{c}\bar{c}$ system, when the channel-coupling includes only three $S$-wave channels, two resonant states are obtained: one with a mass around $7002$ MeV and decay width near $54$ MeV, and another with a mass around $7227$ MeV and a decay width near $66$ MeV.
The former can be regarded as a candidate for the $X(6900)$, and the latter can be considered as a candidate for the $X(7200)$.
Upon adding the $\chi_{c0}\chi_{c2}$, $\chi_{c1}\chi_{c1}$, $\chi_{c1}\chi_{c2}$, $\chi_{c2}\chi_{c2}$ channels, both resonant states still remain. For the $bb\bar{b}\bar{b}$ system, only one resonant state is obtained, regardless of whether the four channels composition of the excited mesons are included or excluded. The mass and width of this resonant state are around $19743$ MeV and $67$ MeV, respectively.
We suggest that future experiments search for the possible resonance state in the invariant mass spectrum of $\Upsilon \Upsilon$ or $\Upsilon \Upsilon(2S)$.
\end{abstract}

\maketitle

\setcounter{totalnumber}{5}

\section{\label{sec:introduction}Introduction}

Following the initial observation of the $X(3872)$ by the Belle Collaboration in 2003~\cite{Belle:2003nnu}, experiments have identified a growing number of analogous exotic (XYZ) states. 
The exploration of exotic states beyond the conventional quark model has emerged as a frontier in particle physics, challenging our understanding of quantum chromodynamics (QCD) and the fundamental nature of strong interactions.

The fully heavy tetraquark states are a relatively pure platform for studying strong interactions, as they are not affected by the long-range light meson exchange mechanism. Experimentally, the exploration of the fully heavy tetraquark states can be traced back to the production of $\Upsilon$-pair reported by the CMS Collaboration in 2017~\cite{CMS:2016liw}.
In 2019, the ANDY Collaboration reported a significant structure at around $18.2$ GeV in Cu+Au collisions~\cite{ANDY:2019bfn}.
However, in 2020, the LHCb Collaboration found no significant results in the $\Upsilon(1S)\mu^{+}\mu^{-}$ invariant mass spectrum at centre-of-mass energies $\sqrt{s}=7, 8$ and $13$ TeV~\cite{LHCb:2018uwm}. Besides, the updated analysis by the CMS Collaboration also revealed no significant excess of events compatible with a narrow resonance~\cite{CMS:2020qwa}.
From a theoretical perspective, some works have been devoted to exploring the possible existence of fully bottomed tetraquark states~\cite{Chen:2016jxd,Bai:2016int,Wang:2021mma,Hughes:2017xie,Eichten:2017ual}.
However, the existence of the fully bottomed tetraquark states remains to be conclusively established in the future through rigorous experimental investigation.

Fortunately, the progress in the experimental exploration of fully charmed tetraquark states is exciting. In 2020, the LHCb Collaboration discovered a new resonant state $X(6900)$ with a global significance exceeding $5 \sigma$ in the $J/\psi J/\psi$ invariant mass spectrum, along with a broad enhancement spanning $6.2$-$6.8$ GeV and a preliminary indication of an additional structure around $7.2$ GeV~\cite{LHCb:2020bwg}. Subsequently, the CMS~\cite{CMS:2023owd} and ATLAS~\cite{ATLAS:2023bft} Collaborations independently confirmed the existence of the $X(6900)$ state. 
Furthermore, the CMS Collaboration reported two additional states, denoted as $X(6600)$ and $X(7200)$, observed in the di-$J/\psi$ invariant mass spectrum with and without interference effects. In the di-$J/\psi$ decay channel, the ATLAS Collaboration not only confirmed the $X(6900)$ but also identified two new structures $X(6400)$ and $X(6600)$. Meanwhile, in the $J/\psi \psi(2S)$ decay channel, the signals of the $X(6900)$ and a new $X(7200)$ state were observed. 

Prior to the experimental observation of $X(6900)$ by the LHCb Collaboration, the fully heavy tetraquark states have been extensively investigated in the literature~\cite{Iwasaki:1975pv,Chao:1980dv,Iwasaki:1976cn,Ader:1981db,Heller:1985cb,Lloyd:2003yc,Barnea:2006sd,Vijande:2009kj,Berezhnoy:2011xn,Anwar:2017toa,Wang:2017jtz,Wu:2016vtq,Karliner:2016zzc,Chen:2018cqz,Debastiani:2017msn,Wang:2019rdo,Richard:2017vry,
Liu:2019zuc,Wang:2018poa}. Following the experimental discovery of these exotic states mentioned above, a wide range of theoretical methods have been applied. Such as the constituent quark model~\cite{Bedolla:2019zwg,Deng:2020iqw,Jin:2020jfc,Lu:2020cns,Zhang:2022qtp,Wu:2024euj,Giron:2020wpx,Yang:2021hrb,Wu:2024ocq,Lu:2025lyu,liu:2020eha,Zhao:2020jvl,Wang:2021kfv,Liu:2021rtn},
QCD sum rules~\cite{Wang:2020ols,Tang:2024zvf,Yang:2020wkh,Agaev:2023wua,Agaev:2023rpj,Agaev:2023gaq,Yu:2022lak,Dong:2022sef,Wu:2022qwd,Chen:2024bpz,Wang:2020dlo,Wan:2020fsk,Wang:2022xja},
potential model~\cite{Mutuk:2021hmi,Lin:2024olg,Huang:2024jin,Zhao:2020nwy},
Bethe-Sapeter framework~\cite{Ke:2021iyh,Li:2021ygk}
and so on~\cite{Karliner:2020dta,Albuquerque:2020hio,Gordillo:2020sgc,Liang:2021fzr,Pal:2021gkr,Ortega:2023pmr,Zhu:2020xni,Celiberto:2025ziy,Wang:2026kcw,Liang:2026rlm,Liu:2025mxv,
Wang:2020wrp,Chao:2020dml,Dong:2020nwy,Zhang:2020hoh,Guo:2020pvt,Cao:2020gul,Zhou:2022xpd}.
In Ref.~\cite{Jin:2020jfc} utilized two constituent quark models (quark delocalization color screening model and chiral quark model) to study the fully heavy tetraquark states. Their computational results show that, for the fully bottomed systems, in the chiral quark model, there are no bound states or resonance states in two structures (the meson-meson structure and diquark-antidiquark structure). However, the quark delocalization color screening model suggests the possible existence of broad resonances with masses ranging from $19.1$ to $19.4$ GeV, and the quantum numbers $J^{P}=0^{+}$, $1^{+}$, and $2^{+}$. For the fully charmed tetraquark states, the $X(6900)$ can be explained as a compact resonance state with $IJ^{P}=00^{+}$. Zhang et al.~\cite{Zhang:2022qtp} employed a nonrelativistic constituent quark model to study the $S$-wave fully heavy tetraquark states. The authors revealed that the $X(6900)$ may not be a ground $cc\bar{c}\bar{c}$ tetraquark state, while $X(6600)$ can be explained as a ground fully charmed tetraquark state with spin-parity $J^{PC}=0^{++}$. Moreover, the states with $bb\bar{b}\bar{b}$ components could be located around $19.2$ GeV. In Ref.~\cite{Wu:2024euj}, the authors found good candidates for the $X(6900)$ and $X(7200)$ in both the $J^{PC}=0^{++}$ and $2^{++}$ systems. Besides, for the fully bottomed tetraquark system, they predicted the existence of some resonant states in the $19.7$-$20.0$ GeV mass range. Additionally, studies on the decays and production of fully heavy tetraquarks can be found in Refs.~\cite{Wang:2023kir,Agaev:2023ruu,Li:2019uch,Chen:2020xwe,Becchi:2020uvq,Chen:2024orv,Yang:2024guo,Feng:2026orq,Wang:2020gmd,Ma:2025ryo,Bai:2024flh,Bai:2024ezn,Feng:2023ghc,
Maciula:2020wri,Goncalves:2021ytq,Feng:2023agq,Wang:2025hex,Celiberto:2025vra,Feng:2020riv,Gong:2020bmg}.

In our previous work~\cite{Wu:2024tif}, we studied the $cc\bar{c}\bar{c}$ system with $J^{PC}=0^{++}$, because this quantum number includes more channels and thus contains richer physical information. Four resonance states, at the energies around $6923$, $6996$, $7060$ and $7159$ MeV, were obtained. However, the decay width of the resonance state at $6923$ MeV is only $10.1$ MeV. Its value is much lower than the experimental measurement. Recently, the CMS Collaboration proposed that the quantum numbers of fully charmed tetraquark states tend to be $J^{PC}=2^{++}$~\cite{CMS:2025fpt}. Given that the experimental data tend to favor the $J^{PC}=2^{++}$ assignment, in this work we systematically investigate the $cc\bar{c}\bar{c}$ states with $J^{PC}=2^{++}$ to search for the candidates of the experimentally reported states in the framework of the chiral quark model. Moreover, we extend our study to the $bb\bar{b}\bar{b}$ system under the same quantum number.

The structure of this paper is as follows. Section II gives a brief description of the quark model and wave functions. Section III is devoted to the numerical results and discussions. The summary is shown in the last section.

\section{MODEL AND WAVE FUNCTIONS}

\subsection{Chiral quark model (ChQM)}
In this work, we use the Chiral quark model (ChQM) to investigate the $QQ\bar{Q}\bar{Q}$ ($Q=c,b$) tetraquark system. The details of the ChQM can be found in Refs.~\cite{Vijande:2004he,Barnea:2006sd}. Here we just present the Hamiltonian of the model.
\begin{eqnarray}
H = \sum_{i=1}^4
\left(m_i+\frac{p_{i}^{2}}{2m_{i}}\right)-T_{CM}+\sum_{j>i=1}^4
\left(V_{CON}+V_{OGE}\right).%\left(V_{ij}^{C}+V_{ij}^{G}+V_{ij}^{\chi} \right), \nonumber\\
\label{Ham}
\end{eqnarray}
Since the quark composition of the system under study consists entirely of heavy quarks, there are no Goldstone and $\sigma$ meson exchange terms, so these two terms are omitted in Eq.(\ref{Ham}). $m_{i}$ $(i=c/b)$ represents the constituent mass of the (charm/bottom) quark. $T_{CM}$ is the kinetic energy of the center of mass.

$V_{CON}$ is the confinement interaction, including central ($V_{CON}^{C}$) and spin-orbit ($V_{CON}^{SO}$) forces, and can be written as:

\begin{subequations}
\label{confinement}
\begin{align}
    V_{CON}^C &= ( -a_{c} r_{ij}^{2}-\Delta) \boldsymbol{\lambda}_i^c \cdot \boldsymbol{\lambda}_j^c\\ \nonumber
    V_{CON}^{SO} &= -\boldsymbol{\lambda}_i^c \cdot \boldsymbol{\lambda}_j^c \frac{a_c}{4m_i^2m_j^2}[ ((m_i^2+m_j^2)(1-2a_s) \\ \nonumber
&+ 4m_im_j(1-a_s))(\vec{S}_{+}\cdot \vec{L}) \\
&+((m_j^2-m_i^2)(1-2a_s))(\vec{S}_{-}\cdot \vec{L}) ]\\ \nonumber
\end{align}
\end{subequations}
where $a_{c}$ is the confinement strength; $\Delta$ represents a global constant fixing the origin of energies; $a_{s}$ governs the ratio between the scalar and vector terms.

$V_{OGE}$ is the one-gluon-exchange interaction, including central ($V_{OGE}^{C}$), spin-orbit ($V_{OGE}^{SO}$) and tensor ($V_{OGE}^{T}$) forces, and can be expressed as follows:

\begin{subequations}
\label{oge}
\begin{align}
    V_{OGE}^C &=\frac{\alpha_{s}}{4} \boldsymbol{\lambda}_i^c \cdot \boldsymbol{\lambda}_{j}^c
\left[\frac{1}{r_{ij}}-\frac{1}{6m_im_jr_0^2}\boldsymbol{\sigma}_i\cdot
\boldsymbol{\sigma}_j \frac{e^{-\frac{r_{ij}}{r_{0}} }}{r_{ij}}\right]   \\ \nonumber
    V_{OGE}^{SO} &= -\frac{1}{16} \frac{\alpha_{s}\boldsymbol{\lambda}_i^c \cdot \boldsymbol{\lambda}_j^c}{4m_i^2m_j^2}[\frac{1}{r_{ij}^3}-\frac{e^{-r_{ij}/r_g(\mu)}}{r_{ij}^3}(1+\frac{r_{ij}}{r_g(\mu)})] \\ \nonumber
    &[ (m_i^2+m_j^2+4m_im_j)(\vec{S}_{+}\cdot \vec{L})\\
    &+(m_j^2-m_i^2)(\vec{S}_{-}\cdot \vec{L}) ]\\ \nonumber
V_{OGE}^{T} &= -\frac{1}{16} \frac{\alpha_{s}\boldsymbol{\lambda}_i^c \cdot \boldsymbol{\lambda}_j^c}{4m_i^2m_j^2}[\frac{1}{r_{ij}^3}-\frac{e^{-r_{ij}/r_g(\mu)}}{r_{ij}}(\frac{1}{r_{ij}^2}\\
&+\frac{1}{3r^2_g(\mu)} +\frac{1}{r_{ij}r_g(\mu)})]S_{ij} \\ \nonumber
\end{align}
\end{subequations}
where $\boldsymbol{\sigma}$ represents the $SU(2)$ Pauli matrices; and $\alpha_{s}$ means the quark-gluon coupling constant (in Table~\ref{modelparameters}, $\alpha_{s_{cc}}$ and $\alpha_{s_{bb}}$ denote the quark-gluon coupling constants for $c\bar{c}$ quark and $b\bar{b}$ quark interactions, respectively); $r_{0}=\hat{r}_{0}/\mu$, scaling with the reduced mass as expected for a Coulombic system; $r_{g}(\mu)=\hat{r}_{g}/\mu$ means a similar behaviour to the scaling of the central term.

The other symbols in the above expressions have their usual meanings.
In this work, the parameters listed in Table~\ref{modelparameters} are all determined by fitting the calculated meson spectrum to the experimental data.
The masses of the mesons calculated for the present work are illustrated in Table~\ref{mesonmass}.

\begin{table}[t]
\begin{center}
\caption{The quark model parameters.\label{modelparameters}}
\begin{tabular}{ccccccccccc}
\hline\hline\noalign{\smallskip}
$m_{c}$(MeV)& $m_{b}$(MeV)& $a_{c}$(MeV) &$\Delta$(MeV)   &$\hat{r}_0$(MeV) &   $\hat{r}_g$(MeV)&\\
        1650&   4977&         98 &         -18.1           &81.0             &   100.6     & \\
 $\alpha_{s_{cc}}$ & $\alpha_{s_{bb}}$ & $a_{s}$ \\
 0.56    & 0.43       & 0.77 \\
\hline\hline
\end{tabular}
\end{center}
\end{table}

\begin{table}[]
\caption{ \label{mesonmass}  Numerical results for the this work ( with a  harmonic form confinement ), the ChQM2( with a  color screening form confinement ). (unit: MeV).}
\begin{tabular}{ccccccc}
\hline\hline\noalign{\smallskip}
    Meson         & ~   This work  ~  & ChQM2\cite{Vijande:2004he}     ~ & EXP.(PDG)\cite{ParticleDataGroup:2024cfk} \\ \hline
    $\eta_{c}$    & ~   2980       ~  &  2990           ~& 2983.9\\
    $\eta_{c}(2S)$& ~   3637       ~  &  3627           ~& 3637.7\\
    $\eta_{c}(3S)$& ~   4132       ~  &   -             ~&              -\\
    $J/\psi$      & ~   3100       ~  &  3097           ~& 3096.9\\
    $\psi(2S)$    & ~   3713       ~  &  3685           ~& 3686.1\\
%    $h_c$         &    3508         &  3507 &   3522 & 3525.37$\pm$0.14\\
    $\chi_{c0}$   & ~   3420       ~  &  3436           ~& 3414.7\\
    $\chi_{c1}$   & ~   3479       ~  &  3494           ~& 3510.6\\
    $\chi_{c2}$   & ~   3523       ~  &  3526           ~& 3556.1\\
    $\eta_{b}$   & ~   9390        ~ &  9454            ~& 9398.7\\
    $\Upsilon$   &  ~  9502        ~ &  9505            ~& 9460.3\\
    $\Upsilon(2S)$   &  ~  9971        ~ &  10013            ~& 10023.2\\
    $\Upsilon(3S)$   &  ~  10289        ~ &  10335            ~& 10355.2\\
    $\chi_{b0}$   & ~   9865        ~ &  9855           ~& 9859.4\\
    $\chi_{b1}$   & ~   9891       ~  &  9875           ~& 9892.7\\
    $\chi_{b2}$   & ~   9908       ~  &  9887           ~& 9912.2\\
\hline\hline
\end{tabular}
\end{table}

\subsection{Wave function}
For the fully heavy system, meson-meson and diquark-antidiquark (denoted as $Q\bar{Q}$-$Q\bar{Q}$ and $QQ$-$\bar{Q}\bar{Q}$ after) structures are considered. The wave function of both structures consists of four parts: orbit, flavor, spin and color wave functions. Considering the spin-orbit coupling effect, the spin and orbital wave functions need to be coupled together, while the flavor and color wave functions are independent.

In Gaussian expansion method (GEM), the radial part of spatial wave function is expanded by Gaussians~\cite{Hiyama:2003cu}:
\begin{subequations}
\label{radialpart}
\begin{align}
\Psi_{L}(\mathbf{r}) & = \sum_{n=1}^{n_{\rm max}} c_{n}\psi^G_{nlm}(\mathbf{r}),\\
\psi^G_{nlm}(\mathbf{r}) & = N_{nl}r^{l}
e^{-\nu_{n}r^2}Y_{lm}(\hat{\mathbf{r}}),
\end{align}
\end{subequations}
where $N_{nl}$ are normalization constants,
\begin{align}
N_{nl}=\left[\frac{2^{l+2}(2\nu_{n})^{l+\frac{3}{2}}}{\sqrt{\pi}(2l+1)}
\right]^\frac{1}{2}.
\end{align}
$c_n$ are the variational parameters, which are determined dynamically. The Gaussian size parameters are selected in accordance with the following geometric progression:
\begin{equation}\label{gaussiansize}
\nu_{n}=\frac{1}{r^2_n}, \quad r_n=r_1a^{n-1}, \quad
a=\left(\frac{r_{n_{\rm max}}}{r_1}\right)^{\frac{1}{n_{\rm
max}-1}}.
\end{equation}
This procedure allows for the optimization of the ranges by utilizing only a small number of Gaussians.

For the spin wave functions, there is no distinction between quark and antiquark. The meson-meson structure possesses the same spin as the diquark-antidiquark structure. The spin wave functions of the subclusters are presented as follows:
\begin{eqnarray}
 & & \chi_{11}^{\sigma}=\alpha\alpha,~~
\chi_{10}^{\sigma}=\frac{1}{\sqrt{2}}(\alpha\beta+\beta\alpha),~~ \nonumber \\
 & & \chi_{1-1}^{\sigma}=\beta\beta,~~\chi_{00}^{\sigma}=\frac{1}{\sqrt{2}}(\alpha\beta-\beta\alpha).
\end{eqnarray}
Once the orbital and spin wave functions have been obtained, the subsequent step is to couple these two types of wave functions. The first step involves obtaining $|\psi_{J_{1}},m_{J_{1}}\rangle$ by coupling the orbital wave function $|\psi_{L_{1}},m_{L_{1}}\rangle$ and spin wave function $|\chi_{S_{1}}^{\sigma},m_{S_{1}}\rangle$ of the first subcluster.
\begin{align}
 |\psi_{J_1,m_{J_1}}\rangle =  |\psi_{L_1,m_{L_1}} \rangle \otimes  |\chi_{S_1,m_{S_1}}^{{\sigma}} \rangle.
\end{align}
Likewise, the orbital wave function $|\psi_{L_{2}},m_{L_{2}}\rangle$ and spin wave function $|\chi_{S_{2}}^{\sigma},m_{S_{2}}\rangle$ of the second subcluster are combined, resulting in the $|\psi_{J_{2}},m_{J_{2}}\rangle$.
\begin{align}
 |\psi_{J_2,m_{J_2}}\rangle =  |\psi_{L_2,m_{L_2}} \rangle \otimes  |\chi_{S_2,m_{S_2}}^{{\sigma}} \rangle.
\end{align}
Then, the wave functions of the two subclusters are coupled to form the $|\psi_{J_{12}},m_{J_{12}}\rangle$.
\begin{align}
 |\psi_{J_{12},m_{J_{12}}}\rangle =  |\psi_{J_1,m_{J_1}}\rangle \otimes  |\psi_{J_2,m_{J_2}}\rangle.
\end{align}
In the final step, the wave function $|\psi_{J_{12}},m_{J_{12}}\rangle$ is coupled with the orbital wave function $|\psi_{L_{3}},m_{L_{3}}\rangle$ between the two subclusters, generating the total orbit-spin wave function $|\psi_{i}^{LS}\rangle$.
\begin{align}
 &|\psi_{i}^{LS}\rangle =|\psi_{J_{12},m_{J_{12}}}\rangle  \otimes  |\psi_{L_3,m_{L_3}} \rangle.  \\
 &i \equiv [J_1,J_2,J_{12},L_3].
 \end{align}
In this work, we focus on the $2^{++}$ system, considering a total of $5$ combinations for $[J_1,J_2,J_{12},L_3]$ under the fixed inter-cluster relative motion in the $S$-wave ($L_{3}=0$), as shown in Table~\ref{coupl}.
\begin{table}[]
\caption{ \label{LScouple}  Different combinations of J-J coupling.\label{coupl}}
\begin{tabular}{cccccccc}
\hline\hline\noalign{\smallskip}
\multicolumn{2}{l}{$L_3=0$} & \multicolumn{2}{c}{$J_{1}=0$} &~~~~&\multicolumn{2}{c}{$J_{2}=2$}& index(i)   \\
                                     && $L_1=1$ & $S_1=1$             && $L_2=1$ & $S_2=1$           & 1       \\\hline
                                     %\cline{3-7}
                                     && \multicolumn{2}{c}{$J_{1}=1$}& &\multicolumn{2}{c}{$J_{2}=1$}& index(i)   \\
                                     && $L_1=0$ & $S_1=1$             && $L_2=0$ & $S_2=1$           & 2       \\
                                     && $L_1=1$ & $S_1=1$             && $L_2=1$ & $S_2=1$           & 3       \\  \hline
                                     && \multicolumn{2}{c}{$J_{1}=1$}& &\multicolumn{2}{c}{$J_{2}=2$}& index(i)   \\
                                     && $L_1=1$ & $S_1=1$             && $L_2=1$ & $S_2=1$           & 4       \\\hline
                                     && \multicolumn{2}{c}{$J_{1}=2$}& &\multicolumn{2}{c}{$J_{2}=2$}& index(i)   \\
                                     && $L_1=1$ & $S_1=1$             && $L_2=1$ & $S_2=1$           & 5       \\
\hline
\end{tabular}
\end{table}

For the flavor wave function, different structures are obtained according to different coupling sequence. For the $Q\bar{Q}$-$Q\bar{Q}$ structure, the coupling sequence is
\begin{align}\label{sec1}
|F_{1}\rangle&= Q_1\bar{Q}_2Q_3\bar{Q}_4
\end{align}
For the $QQ$-$\bar{Q}\bar{Q}$ structure, the coupling sequence is
\begin{align}\label{sec2}
|F_{2}\rangle&= Q_1Q_3\bar{Q}_2\bar{Q}_4
\end{align}

The colorless tetraquark system has four color structures, including $1\otimes1$, $8\otimes8$, $\bar{3}\otimes 3$ and $6\otimes \bar{6}$,
\begin{eqnarray}
|C_{1}\rangle & = & \chi_{1\otimes1}^{m1}=\frac{1}{\sqrt{9}}(\bar{r}r\bar{r}r+\bar{r}r\bar{g}g+\bar{r}r\bar{b}b
   +\bar{g}g\bar{r}r+\bar{g}g\bar{g}g \nonumber \\
  & + & \bar{g}g\bar{b}b+\bar{b}b\bar{r}r+\bar{b}b\bar{g}g+\bar{b}b\bar{b}b), \nonumber \\
|C_{2}\rangle & = & \chi_{8\otimes8}^{m2}=\frac{\sqrt{2}}{12}(3\bar{b}r\bar{r}b+3\bar{g}r\bar{r}g+3\bar{b}g\bar{g}b
   +3\bar{g}b\bar{b}g \nonumber \\
 &+ & 3\bar{r}g\bar{g}r+ 3\bar{r}b\bar{b}r+2\bar{r}r\bar{r}r+2\bar{g}g\bar{g}g+2\bar{b}b\bar{b}b-\bar{r}r\bar{g}g \nonumber \\
&-& \bar{g}g\bar{r}r-\bar{b}b\bar{g}g-\bar{b}b\bar{r}r-\bar{g}g\bar{b}b-\bar{r}r\bar{b}b), \\
|C_{3}\rangle & = & \chi^{d1}_{\bar{3}\otimes 3} =\frac{\sqrt{3}}{6}(rg\bar{r}\bar{g}-rg\bar{g}\bar{r}+gr\bar{g}\bar{r}
    -gr\bar{r}\bar{g}+rb\bar{r}\bar{b}, \nonumber \\
&- & rb\bar{b}\bar{r}+br\bar{b}\bar{r}-br\bar{r}\bar{b}+gb\bar{g}\bar{b}-gb\bar{b}\bar{g}+bg\bar{b}\bar{g}-bg\bar{g}\bar{b}), \nonumber \\
|C_{4}\rangle & = & \chi^{d2}_{6\otimes \bar{6}}=\frac{\sqrt{6}}{12}(2rr\bar{r}\bar{r}+2gg\bar{g}\bar{g}+2bb\bar{b}\bar{b}
    +rg\bar{r}\bar{g} \nonumber \\
&+ &rg\bar{g}\bar{r}+gr\bar{g}\bar{r}+gr\bar{r}\bar{g}+rb\bar{r}\bar{b}+rb\bar{b}\bar{r}+br\bar{b}\bar{r} \nonumber \\
&+ &br\bar{r}\bar{b}+gb\bar{g}\bar{b}+gb\bar{b}\bar{g}+bg\bar{b}\bar{g}+bg\bar{g}\bar{b}).\nonumber
\end{eqnarray}
Where $|C_{1}\rangle$, $|C_{2}\rangle$, $|C_{3}\rangle$ and $|C_{4}\rangle$ represent color singlet-singlet ($1\otimes 1$), color octet-octet ($8\otimes8$), color triplet-antitriplet ($\bar{3}\otimes 3$) and color sextet-antisextet ($6\otimes  \bar{6}$) wave functions, respectively. The state with color wave function $|C_{1}\rangle$ is color-singlet channel, and the one with $|C_{2}\rangle$, $|C_{3}\rangle$ or $|C_{4}\rangle$ is hidden-color channel.

Finally, the total wave function of the tetraquark system is written as:
\begin{equation}
\Psi_{JM_J}^{i,j,k}={\cal A} \psi_{i}^{LS}  F_j C_k,
\end{equation}
Where ${\cal A}$ is the antisymmetrization operator. For the $QQ\bar{Q}\bar{Q}$ system, ${\cal A}=1-P_{13}-P_{24}+P_{13,24}$.
Then, we solve the following Schr\"{o}dinger equation to obtain eigen-energies of the system, with the help of the Rayleigh-Ritz variational
principle.
\begin{equation}
H\Psi_{JM_J}^{i,j,k}=E\Psi_{JM_J}^{i,j,k},
\end{equation}
where $\Psi_{JM_J}^{i,j,k}$ is the wave function of the four-quark states, which is the linear combinations of the above channel wave functions.
\section{Real-scaling method}
\begin{figure}[htp]
  \setlength {\abovecaptionskip} {-0.1cm}
  \centering
  \resizebox{0.50\textwidth}{!}{\includegraphics[width=1.8cm,height=1.6cm]{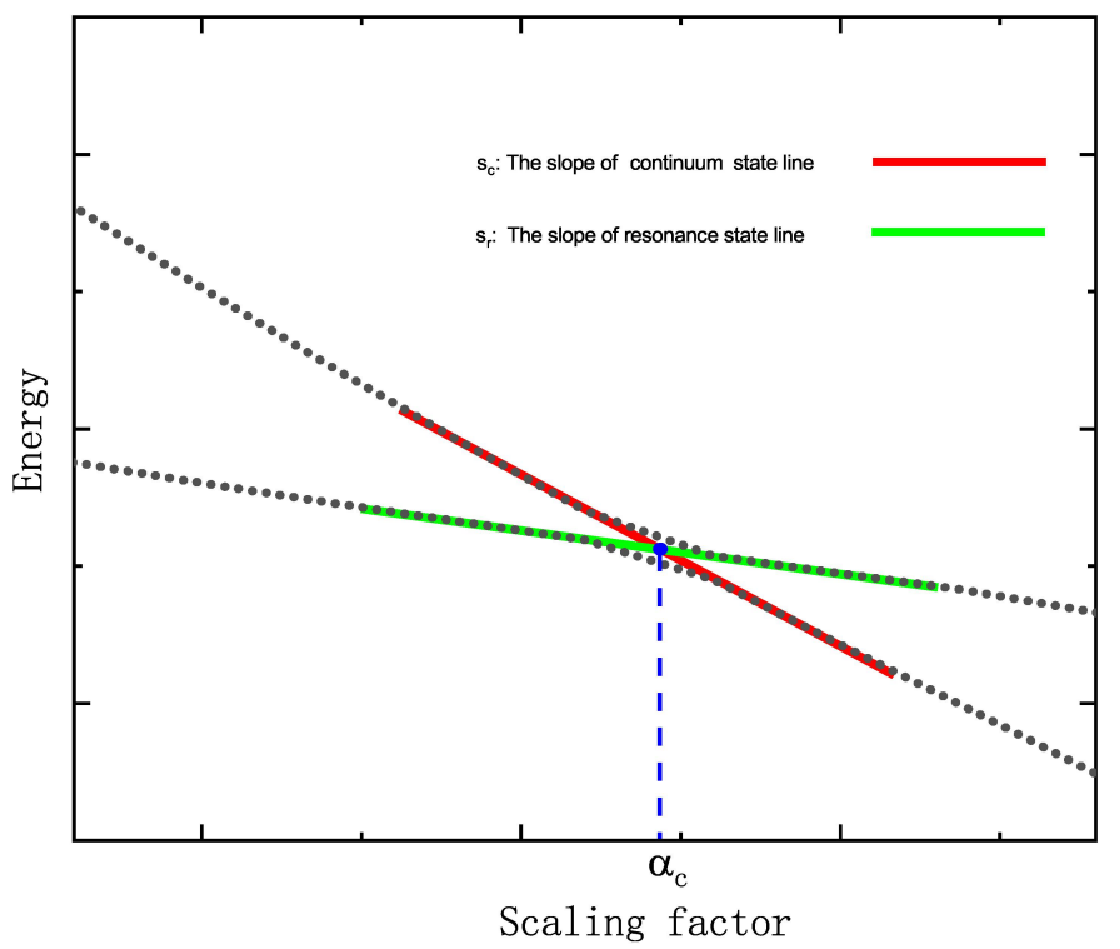}}
  \caption{\label{decaywidth} The schematic energy spectrum in the real-scaling method~\cite{Huang:2023jec}.}
\end{figure}

The real-scaling method, which is also known as the stabilization method~\cite{Simons:1981gbz}, was devised to distinguish the genuine resonances from states with discrete energies during calculations in a finite volume.  This method has been successfully applied to tetraquark systems~\cite{Tan:2020cpu} and pentaquark systems~\cite{Hiyama:2005cf,Hiyama:2018ukv}. In this method, a factor $\alpha$ is utilized for scaling the finite volume. As $\alpha$ increases, the false resonances will decay into the corresponding threshold channels, while the genuine resonances show up as an avoid-crossing configuration (as shown in Fig.~\ref{decaywidth}).
The appearance of the avoid-crossing structure is due to the fact that the energy of one of the scattering states approaches the energy of the resonance states as the scaling factor increases and the coupling becomes stronger. The avoid-crossing structure is a general property of interacting two-level systems. The details have been discussed in Refs.~\cite{Wu:2024tif,Wu:2021rrc,Wu:2023hhk}.

To calculate the decay width of the resonance states, we employ the formula from Ref.~\cite{Simons:1981gbz}:
\begin{equation}\label{formula_RSM}
\Gamma =4 |V(\alpha_c)|\frac{\sqrt{|S_r||S_c|}}{|S_r-S_c|}
\end{equation}
where $V(\alpha_c)$ represents the minimal energy difference between the resonance state and the scattering state; $S_r$ and $S_c$ are the slops of resonance and scattering states, respectively (see Fig.~\ref{decaywidth}).

\section{Result and discussion}
In this work, we investigate the fully charmed and fully bottomed tetraquark systems with quantum number $J^{PC}=2^{++}$ in the framework of ChQM. Two structures: $Q\bar{Q}$-$Q\bar{Q}$ and $QQ$-$\bar{Q}\bar{Q}$ structures, as well as the channel-coupling of the two configurations are considered. In addition to the calculation of
bound states, we also employ the real-scaling method to search for the possible resonant states of the system.
Additionally, we also conduct a comparative study on the impact of including and excluding channels composed of excited mesons ($\chi_{Q0}$, $\chi_{Q1}$, $\chi_{Q2}$, $Q=c$ or $b$) on the results. The notation $\chi_{Q0,1,2}$ $(Q=c, b)$ denotes $P$-wave charmonium/bottomonium states, where the subscript $0,1,2$ indicates the total angular momentum of the $Q\bar{Q}$ $(Q=c, b)$.

\subsection{Fully charmed tetraquark system}

\begin{table}[!t]
\caption{\label{Boundstatecccc} Results of the bound state calculations in the $c\bar{c}c\bar{c}$ system with $J^{PC}=2^{++}$ (unit: MeV).}
\begin{ruledtabular}
\begin{tabular}{cccccc}
Channel             &$|[LS]_i F_j C_k\rangle$  & $E_{th}$           & $E_{sc}$      & $E_{mix1}$  & $E_{mix2}$\\
%$\eta_c \eta_c$     &$|111\rangle$             & $5960$             & $5962$  &5961\\
$J/\psi J/\psi$     &$|211\rangle$             & $6200$             & $6201$   & $6201$ & $6201$\\
$\chi_{c0}\chi_{c2}$&$|111\rangle$             & $6943$             & $6946$  \\
$\chi_{c1}\chi_{c1}$&$|311\rangle$             & $6958$             & $6960$  \\
$\chi_{c1}\chi_{c2}$&$|511\rangle$             & $7002$             & $7005$  \\
$\chi_{c2}\chi_{c2}$&$|412\rangle$             & $7046$             & $7050$  \\
\\
 $[J/\psi]_8[J/\psi]_8$& $|212\rangle$ &                            &6437 & \\
 $[cc]_3^1[\bar{c}\bar{c}]_{\bar{3}}^1$ & $|223\rangle$ &           &6485 & \\

\end{tabular}
\end{ruledtabular}
\end{table}

The energies of the $cc\bar{c}\bar{c}$ tetraquark system for $c\bar{c}$-$c\bar{c}$ and $cc$-$\bar{c}\bar{c}$ structures, as well as the channel coupling of these two structure are listed in Table~\ref{Boundstatecccc}. Where $E_{th}$ represents the theoretical threshold; $E_{sc}$ is the energy of every single channel; 
$E_{mix1}$ is the lowest energy of the system by coupling three $S$-wave channels ($J/\psi J/\psi$, $[J/\psi]_{8} [J/\psi]_{8}$,$[cc]_{3}^{1}[\bar{c}\bar{c}]_{3}^{1}$); and $E_{mix2}$ means the lowest energy of the system by coupling all seven channels of both configurations.

For the $J^{PC}=2^{++}$ system, we analyze seven channels, including five color-singlet channels and two color-exotic channels (an octet-octet channel and a triplet-antitriplet channel).
It can be seen from Table~\ref{Boundstatecccc} that, the energies of the five color-singlet channels are all higher than the corresponding thresholds, which means none of them will form a bound state. Besides, the energies of both color-exotic channels are above the lowest color-singlet ($J/\psi J/\psi$) channel.
When only considering the coupling of three $S$-wave channels, the energy $E_{mix1}$ is $6201$ MeV, which is higher than the lowest $J/\psi J/\psi$ threshold. Then, after adding four channels ($\chi_{c0}\chi_{c2}$, $\chi_{c1}\chi_{c1}$, $\chi_{c1}\chi_{c2}$, $\chi_{c2}\chi_{c2}$) composed of excited mesons and coupling all seven channels, the lowest energy $E_{mix2}$ is almost unchanged. This indicates that the newly added channels, after channel coupling, cannot help too much to form a bound state below the lowest threshold. This is mainly because of the large mass gap between these four channels and the three $S$-wave channels.
%Therefore, there is no bound state below the lowest threshold (6200 MeV) in this system.
Besides, the numerical results for contributions from each term in the Hamiltonian has been shown in Table~\ref{2c}. It can be seen from the Table~\ref{2c} that the spin-orbit coupling effect exists only in channels composed of excited mesons. This is because for channels formed by ground-state mesons, the orbital angular momentum is
$L=0$, so the spin-orbit coupling does not work. For channels composed of excited mesons, except for the$V_{OGE}^{SO}$ term in the $\chi_{c0}\chi_{c0}$ channel which is repulsive, the other channels provide attractive potentials. The $V_{OGE}^{C}$ term in all channels provides most of the attractive potential, but it is not sufficient to lower the energy below the corresponding thresholds. After channel coupling, since no bound state is formed, the respective contributions of each term for $E_{mix1}$ and $E_{mix2}$ are basically identical to those in the individual $J/\psi J/\psi$ channel.
Therefore, there is no bound state below the lowest threshold (6200 MeV) in this system.

\begin{table*}[!htb]
\caption{\label{2c} The numerical results for contributions from each term in the Hamiltonian with $J^{PC}=2^{++}$ in the $cc\bar{c}\bar{c}$ system (Unit:MeV).}
\begin{ruledtabular}
\begin{tabular}{ccccccccccccc}
Channel     & $J/\psi J/\psi$  & $\chi_{c0}\chi_{c2}$ & $\chi_{c1}\chi_{c1}$ & $\chi_{c1}\chi_{c2}$ & $\chi_{c2}\chi_{c2}$ & $[J/\psi]_{8} [J/\psi]_{8}$ &  $[cc]_3^1[\bar{c}\bar{c}]_{\bar{3}}^1$  & $E_{mix1}$  & $E_{mix2}$\\ %&total\\
mass        & $6600$            & $6600$              & $6600$        & $6600$          & $6600$        & $6600$      & $6600$       & $6600$   & $6600$             \\
$E_{k}$     & $682$             & $777$               & $754$            &$709$          &$664$            &$630$         &$670$     & $682$    & $682$  \\%&$100\%$\\
$V^{C}_{CON}$ &$-15$          & $227$                  & $230$           &$255$         &$280$         &$56$          &$62$     &$-15$  &$-15$\\%&$100\%$    \\
$V^{SO}_{CON}$ & $0$         & $1$                    & $1$            &$0$           &$-2$            &$0$           &$0$      &$0$    &$0$  \\%&$100\%$  \\
$V^{C}_{OGE}$  &$-1066$       & $-586$                & $-579$          &$-559$         &$-538$        &$-849$        &$-847$   &$-1066$    &$-1066$  \\%&$100\%$  \\
$V^{SO}_{OGE}$ & $0$           & $-52$                & $-61$            &$-6$             &$49$           &$0$            &$0$ & $0$   & $0$ \\
$V^{T}_{OGE}$  &$0$           & $-21$                   & $15$          &$6$             &$-3$        &$0$           &$0$   &$0$ &$0$ \\
total       &$6201$         &$6946$                     &$6960$               &$7005$            &$7050$            &$6437$       &$6485$  &$6201$  &$6201$\\
\end{tabular}
\end{ruledtabular}
\end{table*}

\begin{figure}[ht]
\begin{center}
\epsfxsize=3.39in \epsfbox{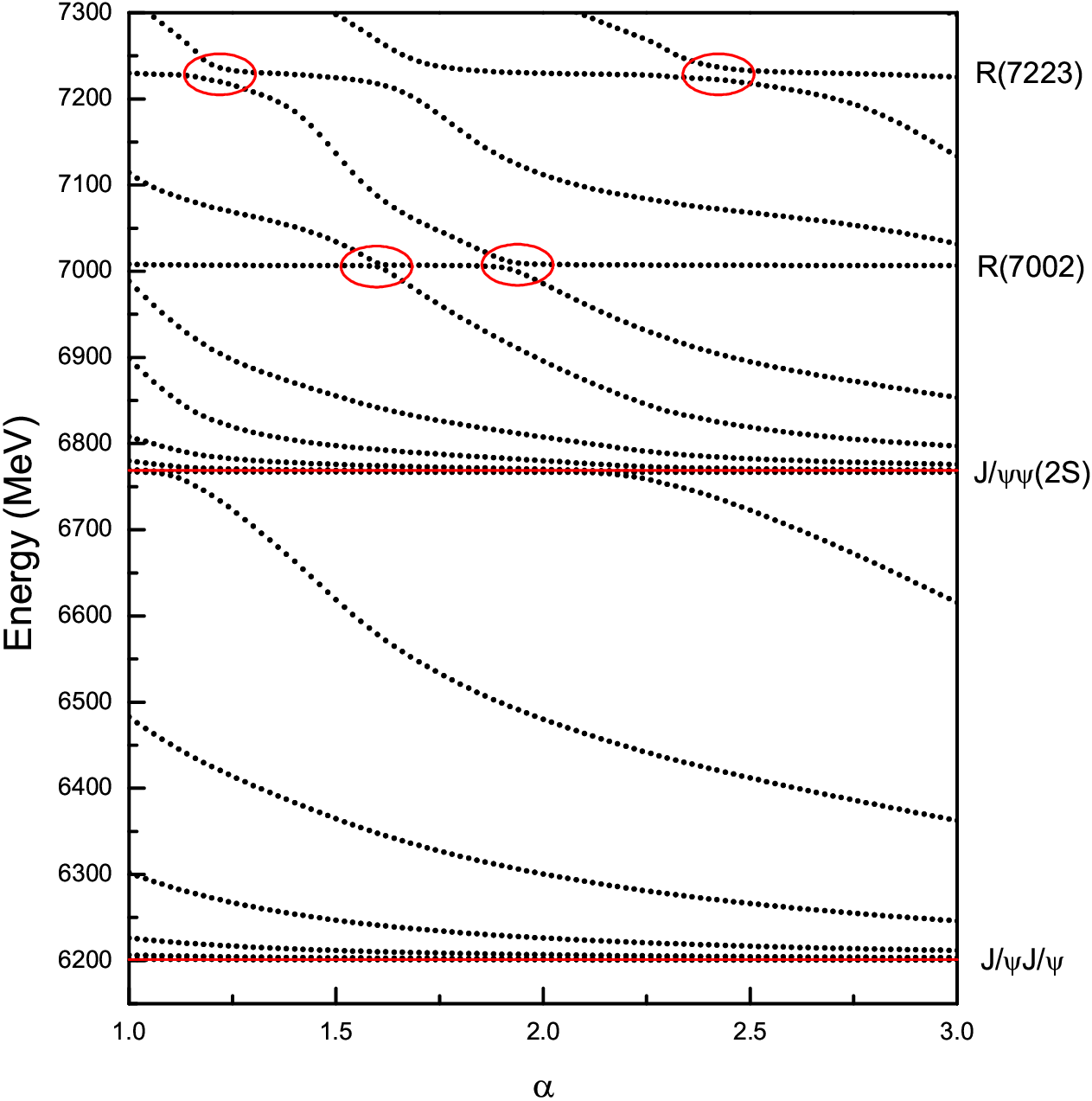} \vspace{-0.2in}

\caption{\label{resonc3}The stabilization plots of the energies including three channels ($J/\psi J/\psi$, $[J/\psi]_{8} [J/\psi]_{8}$,$[cc]_{3}^{1}[\bar{c}\bar{c}]_{3}^{1}$) in the $cc\bar{c}\bar{c}$ system with $J^{PC}=2^{++}$.}
\end{center}
\end{figure}

\begin{figure}[ht]
\begin{center}
\epsfxsize=3.39in \epsfbox{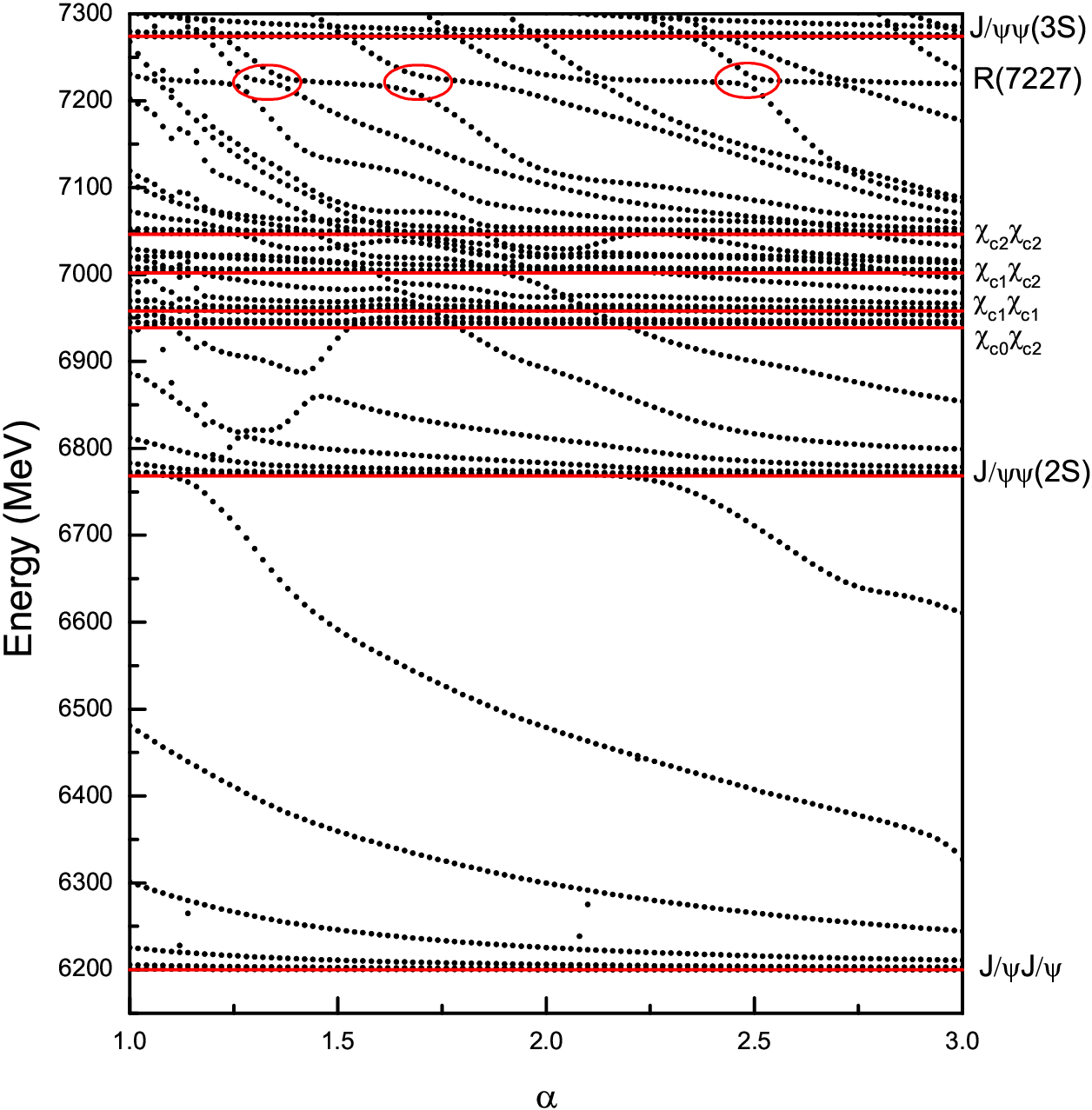} \vspace{-0.2in}

\caption{\label{resonc7}The stabilization plots of the energies including seven channels in the $cc\bar{c}\bar{c}$ system with $J^{PC}=2^{++}$.}
\end{center}
\end{figure}

For the color-excited structures, the phenomenon of color confinement prevents them from dissociating directly. These structures thus have the potential to form resonant states.
In order to identify the genuine resonance state in this system, the real-scaling method is employed here, and the corresponding results are presented in Figs.~\ref{resonc3} and \ref{resonc7}. It should be noted that if there is a large number of coupled channels, the avoid-crossing structures will also be formed because of the different rates of the scattering channel descending to the threshold line. Thus, it is necessary to calculate the components at the resonant state to check whether it is a genuine resonant state.

In Fig.~\ref{resonc3}, the thresholds of $J/\psi J/\psi$ and $J/\psi \psi(2S)$ are marked by red horizontal lines. Near the energies of $7002$ and $7223$ MeV, two avoid-crossing structures are marked by red circles (denoted as $R(7002)$ and $R(7223)$), respectively.
In addition, the color structure proportions at $R(7002)$ and $R(7223)$ are $90\%$ ($45\%$ of the $\bar{3}\otimes3$ configuration) and $92\%$ ($71\%$ of the $\bar{3}\otimes3$ configuration) respectively. Therefore, we consider that these two structures are genuine resonance states. To calculate the decay width of the resonance state, we employ Eq.~(\ref{formula_RSM}) as described in Section III. Then the decay widths of $R(7002)$ and $R(7223)$ that we obtain are $54$ and $66$ MeV, respectively.

From Fig.~\ref{resonc7}, we can see that the red horizontal lines from bottom to top represent the thresholds of $J/\psi J/\psi$, $J/\psi \psi(2S)$, $\chi_{c0}\chi_{c2}$, $\chi_{c1}\chi_{c1}$, $\chi_{c1}\chi_{c2}$, $\chi_{c2}\chi_{c2}$,$J/\psi \psi(3S)$, respectively. At the energy around $7227$ MeV, three avoid-crossing structures are marked by red circles. This means there may be a resonance state near this energy, and it is denoted as $R(7227)$. 
The calculation result shows that the component at the energy of $7227$ MeV is predominantly color structures, accounting for $82\%$ ($51\%$ of the $\bar{3}\otimes3$ configuration). Therefore, the resonant state $R(7227)$ is a genuine resonant state rather than a false resonant state formed by scattering states.
Besides, the decay width of $R(7227)$ that we obtain is $61$ MeV.

\begin{figure}[ht]
\begin{center}
\epsfxsize=3.39in \epsfbox{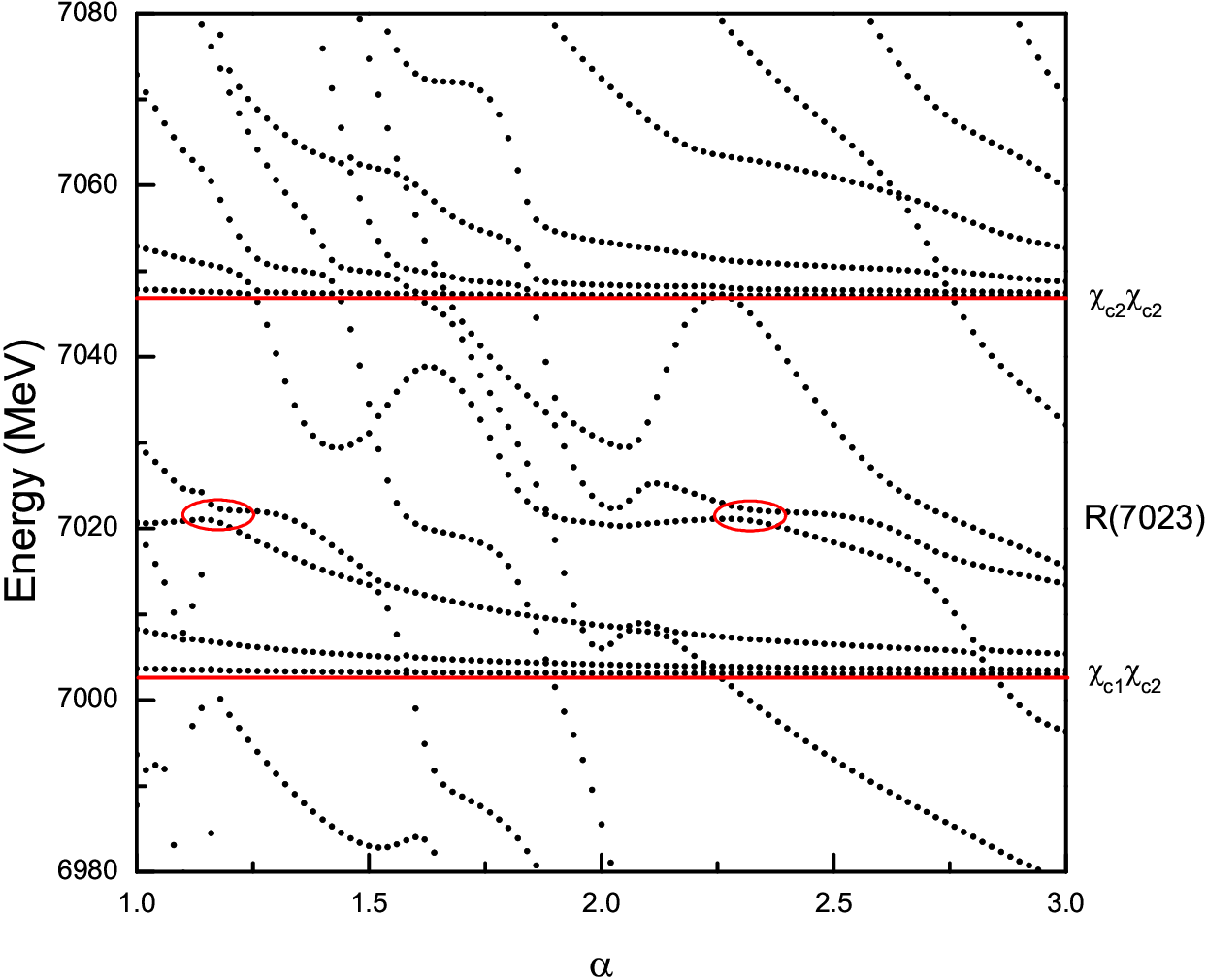} \vspace{-0.2in}

\caption{\label{resonc71}The stabilization plots of the energies including seven channels in the $cc\bar{c}\bar{c}$ system with $J^{PC}=2^{++}$ (energy range from $6980$ to $7080$ MeV).}
\end{center}
\end{figure}

In Fig.~\ref{resonc7}, the energies near $6980-7080$ MeV are too dense to directly judge the presence of the resonance state. To better reveal the avoid-crossing structure in this region, we zoom in on this energy range to produce Fig.~\ref{resonc71}. It clearly shows two avoid-crossing structures (marked by red circles) near the energy of $7023$ MeV (marked as $R(7023)$) in Fig.~\ref{resonc71}. Moreover, the main component of $R(7023)$ is color structures, accounting for $65\%$ ($42\%$ of the $\bar{3}\otimes3$ configuration). Therefore, it is a genuine resonance state. From Eq.(\ref{formula_RSM}), its width is calculated to be $43$ MeV.

The above results indicate that the two resonances $R(7223)$ and $R(7002)$ in Fig.~\ref{resonc7} persist after the inclusion of channels consisting of excited mesons, but their energies are pushed upward due to the interaction between the new channels and the original resonances. Therefore, we conclude that $R(7223)$ and $R(7227)$ are the same resonance, and likewise $R(7002)$ and $R(7023)$ are the same resonance. We retain the former designation for each, namely $R(7223)$ and $R(7002)$.
The resonance state $R(7002)$ may serve as a candidates for the $X(6900)$, while $R(7223)$ is a good candidate for the $X(7200)$.
In addition, from the perspective of their structure proportions, the resonance $R(7002)$ tends to have a molecular structure, while $R(7223)$ is more inclined toward a compact tetraquark structure.
In Ref.~\cite{Wu:2024euj}, the authors consider that for the $cc\bar{c}\bar{c}$ system with $J^{PC}=2^{++}$, the lower resonant state with mass $M\approx 7000$ MeV and width $\Gamma\approx75$ MeV may serve as the candidate for the $X(6900)$. Meanwhile, a higher resonant state with mass $M\approx 7200$ MeV and width $\Gamma\approx50$ MeV could be the candidate for the $X(7200)$. This is consistent with the conclusion we obtained herein.

\subsection{Fully bottomed tetraquark system}
The energies of the $bb\bar{b}\bar{b}$ tetraquark system with $J^{PC}=2^{++}$ are listed in Table~\ref{Boundstatebbbb}.
\begin{table}[!t]
\caption{\label{Boundstatebbbb} Results of the bound state calculations in the $b\bar{b}b\bar{b}$ system with $J^{PC}=2^{++}$ (unit: MeV).}
\begin{ruledtabular}
\begin{tabular}{cccccc}
Channel             &$|[LS]_i F_j C_k\rangle$  & $E_{th}$           & $E_{sc}$      & $E_{mix1}$  & $E_{mix2}$\\
%$\eta_c \eta_c$     &$|111\rangle$             & $5960$             & $5962$  &5961\\
$\Upsilon \Upsilon$     &$|211\rangle$             & $19904$             & $19005$   & $19005$  & $19005$ \\
$\chi_{b0}\chi_{b2}$&$|111\rangle$             & $19773$             & $19774$  \\
$\chi_{b1}\chi_{b1}$&$|311\rangle$             & $19782$             & $19782$  \\
$\chi_{b1}\chi_{b2}$&$|511\rangle$             & $19799$             & $19801$  \\
$\chi_{b2}\chi_{b2}$&$|412\rangle$             & $19816$             & $19818$  \\
\\
 $[\Upsilon]_8[\Upsilon]_8$& $|212\rangle$ &                            &19284 & \\
 $[bb]_3^1[\bar{b}\bar{b}]_{\bar{3}}^1$ & $|223\rangle$ &           &19330 & \\

\end{tabular}
\end{ruledtabular}
\end{table}
From numerical results in Table~\ref{Boundstatebbbb} we can see that, for the $b\bar{b}$-$b\bar{b}$ structures, there is no bound state below the corresponding thresholds. Besides, the energy of $bb$-$\bar{b}\bar{b}$ is several hundred MeV higher than the lowest threshold ($18998$ MeV).
When considering the channel-coupling calculations, regardless of whether channels composed of excited mesons ($\chi_{b0}\chi_{b2}$, $\chi_{b1}\chi_{b1}$, $\chi_{b1}\chi_{b2}$, $\chi_{b2}\chi_{b2}$) are added or not, there is no bound state below the lowest threshold.
Besides, the numerical results for contributions from each term in the Hamiltonian has been shown in Table~\ref{1b}. Similar to the cccc system, the $V_{OGE}^{C}$ term provides the majority of the attractive potential, yet it is insufficient to yield a bound state.

\begin{table*}[!htb]
\caption{\label{1b} The numerical results for contributions from each term in the Hamiltonian with $J^{PC}=2^{++}$ in the $bb\bar{b}\bar{b}$ system (Unit:MeV).}
\begin{ruledtabular}
\begin{tabular}{ccccccccccccc}
Channel     & $\Upsilon\Upsilon$  & $\chi_{b0}\chi_{b2}$ & $\chi_{b1}\chi_{b1}$ & $\chi_{b1}\chi_{b2}$ & $\chi_{b2}\chi_{b2}$ & $[\Upsilon]_{8} [\Upsilon]_{8}$ &  $[bb]_3^1[\bar{b}\bar{b}]_{\bar{3}}^1$   & $E_{mix1}$  & $E_{mix2}$\\ %&total\\
mass        & $19908$            & $19908$           & $19908$        & $19908$          & $19908$        & $19908$      & $19908$  & $19908$   & $19908$                   \\
$E_{k}$     & $803$             & $563$               & $545$            &$520$          &$494$            &$651$         &$642$     & $803$    & $803$   \\
$V^{C}_{CON}$ &$-138$          & $2$                  & $4$           &$12$         &$20$         &$-111$          &$-100$   &$-138$    &$-138$\\
$V^{SO}_{CON}$ & $0$         & $\approx0$                    & $0.1$            &$0$           &$-0.1$            &$0$           &$0$     & $0$  & $0$    \\
$V^{C}_{OGE}$  &$-1568$       & $-668$                & $-658$          &$-640$         &$-623$        &$-1164$        &$-1120$   &$-1568$  &$-1568$  \\
$V^{SO}_{OGE}$ & $0$           & $-21$                & $-24$            &$-2$             &$20$           &$0$        &$0$   & $0$  & $0$\\
$V^{T}_{OGE}$  &$0$           & $-10$                   & $7$          &$3$             &$-1$        &$0$           &$0$   & $0$     & $0$\\
total       &$19005$         &$19774$                     &$19782$               &$19801$            &$19818$            &$19284$       &$19330$    &$19005$    &$19005$
\end{tabular}
\end{ruledtabular}
\end{table*}

\begin{figure}[ht]
\begin{center}
\epsfxsize=3.39in \epsfbox{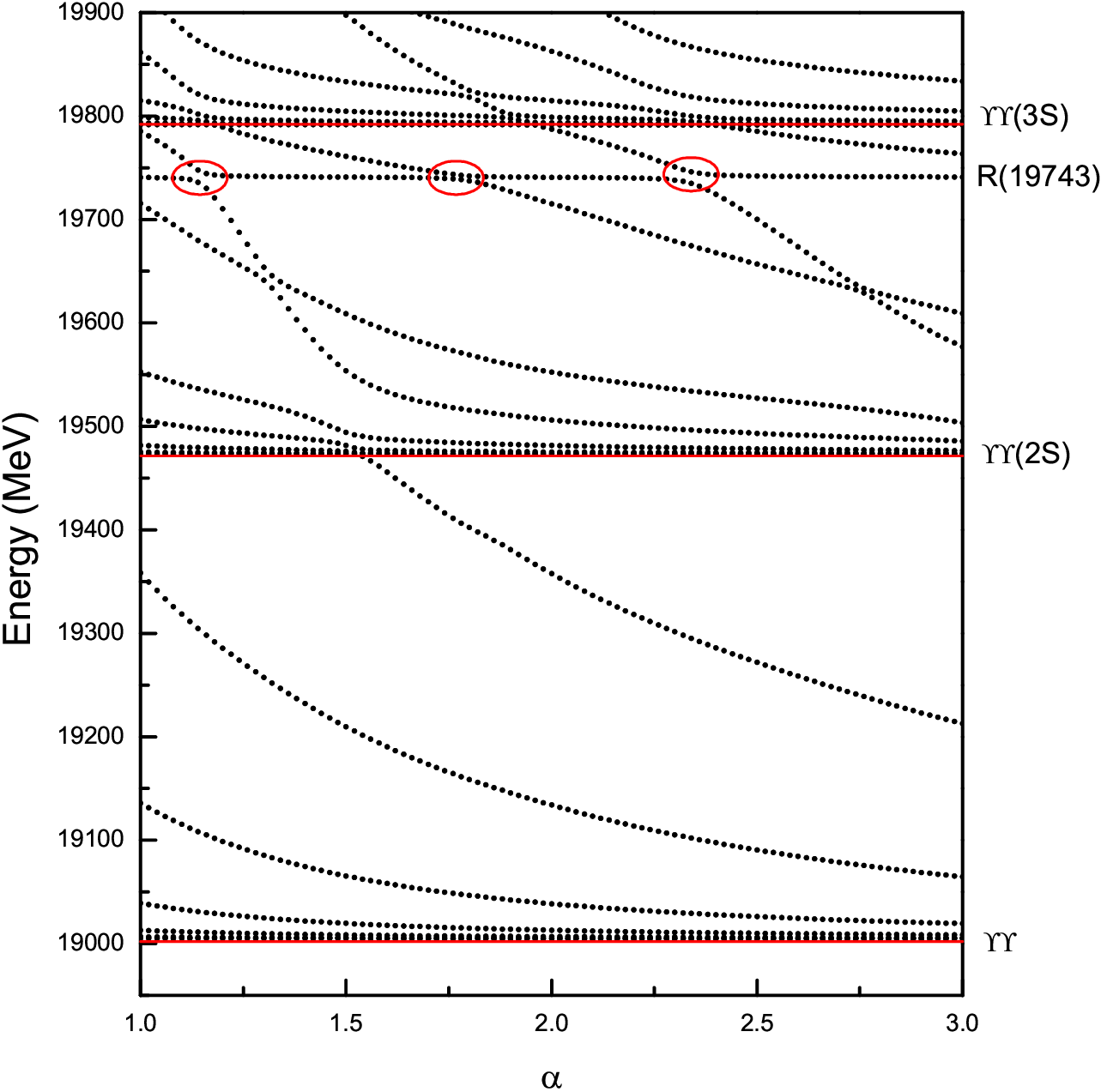} \vspace{-0.2in}

\caption{\label{resonb3}The stabilization plots of the energies including three channels ($\Upsilon \Upsilon$, $[\Upsilon]_{8} [\Upsilon]_{8}$,$[bb]_{3}^{1}[\bar{b}\bar{b}]_{3}^{1}$) in the $bb\bar{b}\bar{b}$ system with $J^{PC}=2^{++}$.}
\end{center}
\end{figure}

\begin{figure}[ht]
\begin{center}
\epsfxsize=3.39in \epsfbox{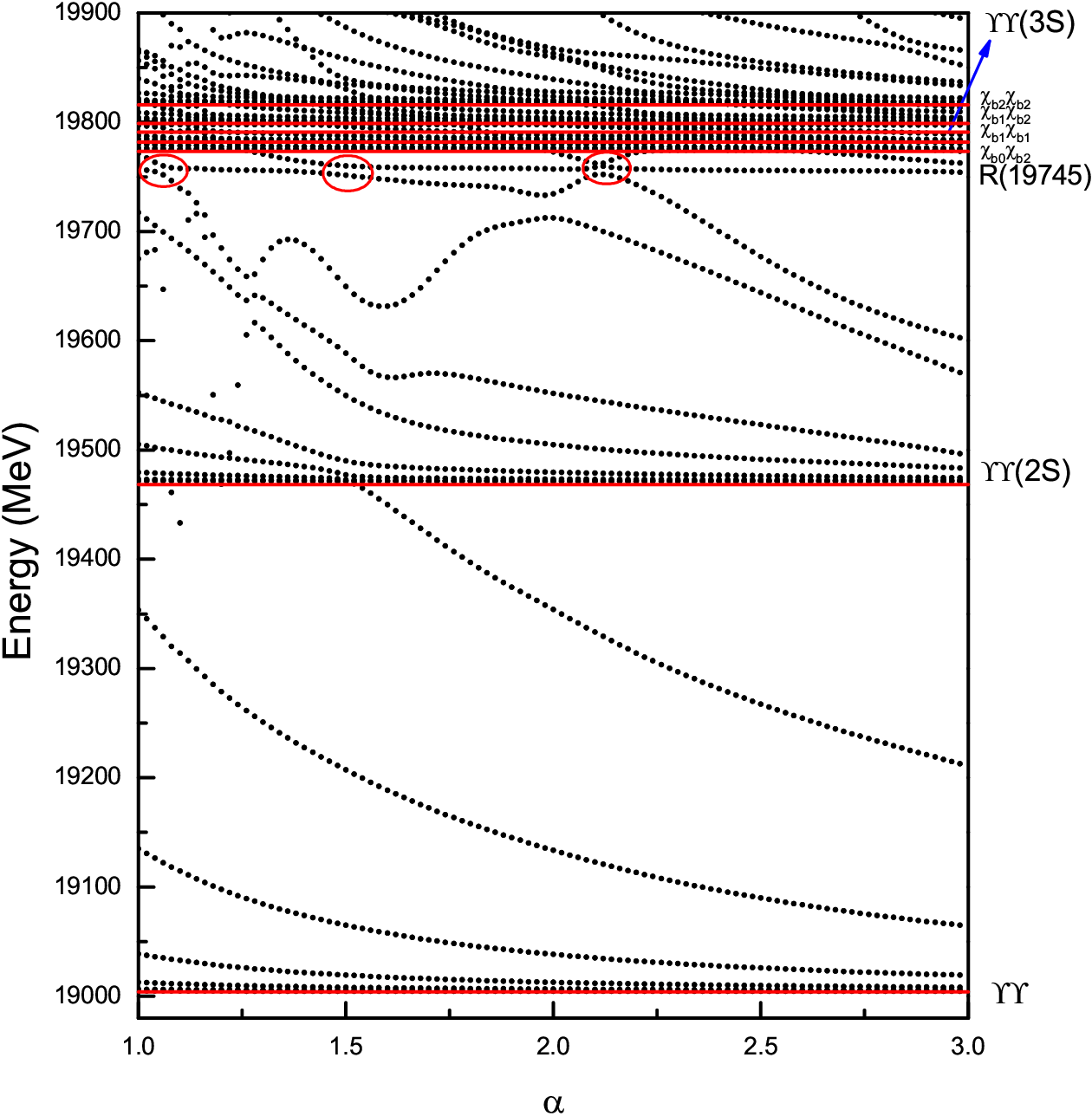} \vspace{-0.2in}

\caption{\label{resonb7}The stabilization plots of the energies including seven channels in the $bb\bar{b}\bar{b}$ system with $J^{PC}=2^{++}$.}
\end{center}
\end{figure}

Furthermore, the real-scaling method is also utilized to search for the genuine resonance states in this system. The result is shown in Fig.~\ref{resonb3} and~\ref{resonb7}.
In Fig.~\ref{resonb3}, the red horizontal lines from bottom to top represent the thresholds of $\Upsilon \Upsilon$, $\Upsilon \Upsilon(2S)$ and $\Upsilon \Upsilon(3S)$, respectively. Near the energy of $19743$ MeV, the avoid-crossing structures are marked by red circles (denoted as $R(19743)$). Moreover, the color structure proportion at the energy of $19743$ MeV is $88\%$ ($66\%$ of the $\bar{3}\otimes3$ configuration). Therefore, the state $R(19743)$ is a genuine resonance state. The decay width of $R(19743)$ is obtained as $68$ MeV.
Besides, at the energy around $19645$ MeV, the avoid-crossing structures are also formed. However, its composition is mainly scattering channels, accounting for $64\%$. Therefore, we believe that this type of avoid-crossing structure is caused by the different rate of the scattering channels descending to the threshold line. In this way, it is a false resonant state at the energy around $19645$ MeV.

In Fig.~\ref{resonb7}, the red horizontal lines from bottom to top represent the thresholds of $\Upsilon \Upsilon$, $\Upsilon \Upsilon(2S)$, $\chi_{b0}\chi_{b2}$, $\chi_{b1}\chi_{b1}$, $\Upsilon \Upsilon(3S)$, $\chi_{b1}\chi_{b2}$, $\chi_{b2}\chi_{b2}$, respectively.
At the energy of $19745$ MeV, two avoid-crossing structures are marked by red circles. There may be a resonant state here, and it is named $R(19745)$. %Like the resonance state $R(7227)$ in $cc\bar{c}\bar{c}$ system, the $R(19745)$ state should also be checked whether it is a genuine resonant state.
The $79\%$ ($64\%$ of the $\bar{3}\otimes3$ configuration) color structure proportion indicates that $R(19745)$ is a genuine resonant state, and its decay width is $64$ MeV. The newly added channels $\chi_{b0}\chi_{b2}$, $\chi_{b1}\chi_{b1}$, $\chi_{b1}\chi_{b2}$ and $\chi_{b2}\chi_{b2}$ do not make $R(19743)$ in Fig.~\ref{resonb3} vanish, as the energies of these channels are all above that of $R(19743)$. Thus it can not decay through these channels.
In addition, from the perspective of their structure proportions, the resonance $R(19743)$ is more inclined toward a compact tetraquark structure.

\begin{figure}[ht]
\begin{center}
\epsfxsize=3.39in \epsfbox{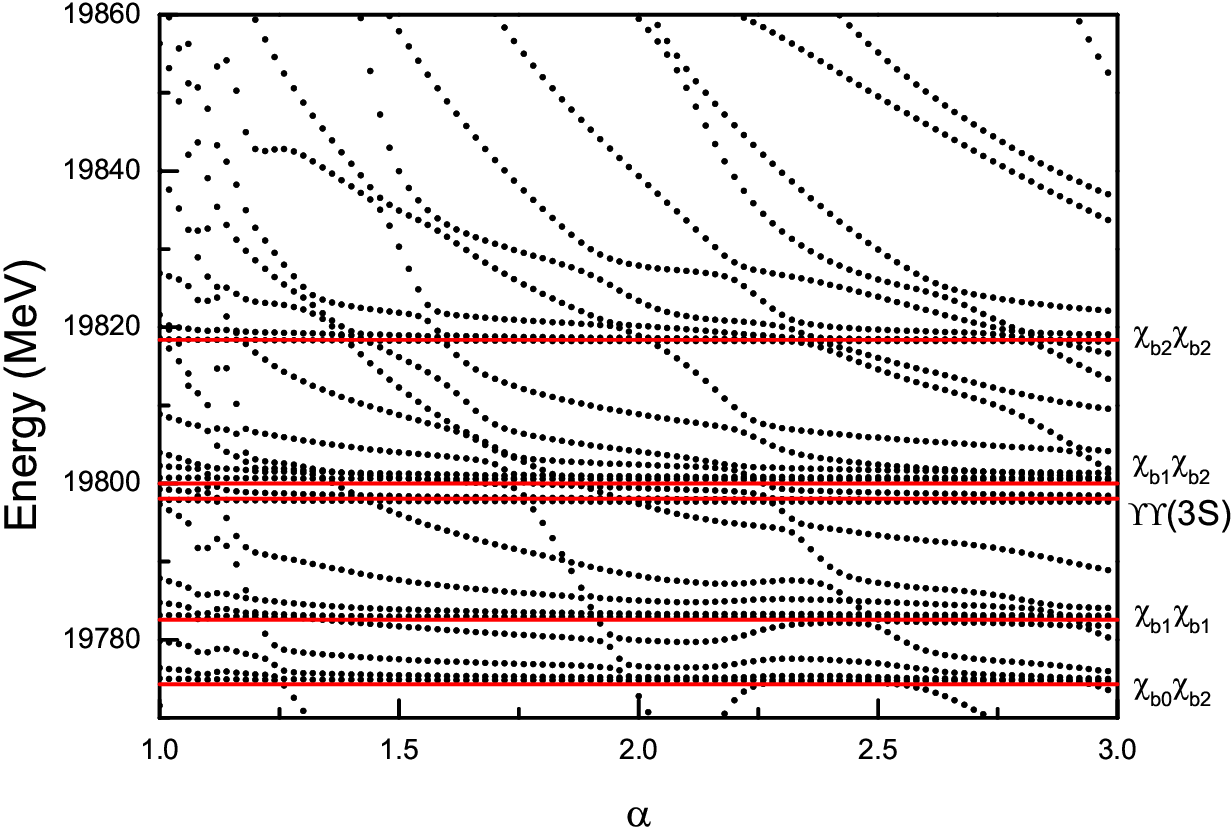} \vspace{-0.2in}

\caption{\label{resonb71}The stabilization plots of the energies including seven channels in the $bb\bar{b}\bar{b}$ system with $J^{PC}=2^{++}$(energy range from $19770$ to $19860$).}
\end{center}
\end{figure}

Using the same handling as for the $cc\bar{c}\bar{c}$ system, we performed a magnified observation of the energy region from $19770$ to $19860$ MeV in Fig.~\ref{resonb7} (as shown in Fig.~\ref{resonb71}). The results show that the avoid-crossing structures are dominated by threshold channel configurations, and hence no resonance states exist.

\section{summary}
In this work, we systematically investigate the fully charmed and fully bottomed tetraquark systems with $J^{PC}=2^{++}$ in the framework of the ChQM. Two structures, $Q\bar{Q}-Q\bar{Q}$ and $QQ-\bar{Q}\bar{Q}$, as well as the coupling of these two configurations are considered.
The dynamical bound-state calculation is carried out to search for any bound state in the fully heavy systems. Meanwhile, a stabilization calculation (real-scaling method) is carried out to find any resonance state.

The bound-state calculation shows that there is no bound state lower than the lowest threshold for the $cc\bar{c}\bar{c}$ and $bb\bar{b}\bar{b}$ systems with $J^{PC}=2^{++}$.
However, in the $cc\bar{c}\bar{c}$ system, by using the real-scaling method, two resonant states $R(7002)$ and $R(7227)$ are obtained. The decay widths of these two resonant states are $54$ MeV and $66$ MeV, respectively. $R(7002)$ can be regarded as the candidate for the $X(6900)$, while $R(7227)$ corresponds to $X(7200)$.
Our results are consistent with Ref.~\cite{Wu:2024euj} in the $J^{PC}=2^{++}$ system, where the resonant state with mass $M\approx 7000$ MeV and width $\Gamma\approx75$ MeV may serve as a candidate for the $X(6900)$. Meanwhile, a higher resonant state with mass $M\approx 7200$ MeV and width $\Gamma\approx50$ MeV could be a candidate for the $X(7200)$.

In the $bb\bar{b}\bar{b}$ system, one resonance state is obtained at the energy around $19743$ MeV. 
The decay width of the resonant state is around $68$ MeV. Moreover, in Ref.~\cite{Wu:2024euj}, the authors also obtained a resonant state with a mass of $19788$ MeV and a width of $60$ MeV. We propose to experimentally search for this possible resonant states with a larger amount of data on the invariant mass spectrum of $\Upsilon \Upsilon$ or $\Upsilon \Upsilon(2S)$.

\acknowledgments{
This work is supported partly by the National Natural Science Foundation of China under Grant Nos. 12575088, 11675080, 11775118, 11535005 and 12305087, and the Funding for School-Level Research Projects of Yancheng Institute of Technology No.xjr2025010 and 2022039, the Start-up Funds of Nanjing Normal University under Grant No. 184080H201B20, and School-Level Research Projects of Henan normal university No. 20240304, and Changzhou Sci\&Tech Program under Grant No. 20260098.
Y. T. is supported by the Qinglan Project of Jiangsu Province of China.}

\end{document}